\documentclass[10pt,conference]{IEEEtran}

\newtheorem{theorem}{Theorem}
\newtheorem{lemma}{Lemma}
\usepackage{epsfig}
\usepackage{graphicx}
\usepackage{amsmath}
\usepackage{amssymb}
\def\mX{\mathcal{X}}
\def\mY{\mathcal{Y}}
\def\mP{\mathcal{P}}
\def\mC{\mathcal{C}}
\def\mD{\mathcal{D}}
\def\mW{\mathcal{W}}

\def\x{\vec{\mathbf{x}}}
\def\y{\vec{\mathbf{y}}}
\def\s{\mathbf{s}}
\def\1{\mathbf{1}}

\def\B{\bigg}
\def\h{\Bigg}
\def\l{\left}
\def\r{\right}

\begin{document}

\title{The source coding game with a cheating switcher}

\author{
\authorblockN{Hari Palaiyanur}
\authorblockA{hpalaiya@eecs.berkeley.edu} \and
\authorblockN{Cheng Chang}
\authorblockA{cchang@eecs.berkeley.edu \\\\
Department of Electrical Engineering and Computer Sciences \\
University of California at Berkeley \\Berkeley, CA 94720} \and
\authorblockN{Anant Sahai}
\authorblockA{sahai@eecs.berkeley.edu}}
\maketitle
\begin{abstract}  Berger's paper `The Source Coding Game', \emph{IEEE Trans. Inform. Theory}, 1971, considers
the problem of finding the rate-distortion function for an adversarial source comprised of
multiple known IID sources. The adversary, called the `switcher', was allowed only causal
access to the source realizations and the rate-distortion function was obtained through the use
of a type covering lemma. In this paper, the rate-distortion function of the adversarial source
is described, under the assumption that the switcher has non-causal access to all source
realizations. The proof utilizes the type covering lemma and simple conditional, random
`switching' rules. The rate-distortion function is once again the maximization of the $R(D)$
function for a region of attainable IID distributions.
\end{abstract}

\section{Introduction}

    The rate distortion function, $R(D)$, specifies the number of codewords, on an exponential scale,
needed to represent a source to within a distortion $D$. Shannon \cite{ShannonRateDistortion}
showed that for an additive distortion function $d$ and a known discrete source that produces
independent and identically distributed (IID) letters according to a distribution $p$,
    \begin{equation}
        R(D) = R_p(D) \triangleq \min_{W:\sum_{x,y} p(x)W(y|x)d(x,y) \leq D}I(p,W)
    \end{equation}
where $I(p,W)$ is the mutual information for an input distribution $p$ and probability
transition matrix $W$.

Sakrison \cite{SakrisonCompoundSources} studied the rate distortion function for the class of
{\em compound} sources. That is, the source is assumed to come from a known set of
distributions and is fixed for all time. If $G$ is the set of possible sources, Sakrison showed
that planning for the worst case source is both necessary and sufficient in the discrete
memoryless source case. Hence, for compound sources,
    \begin{equation}
        R(D) = \max_{p \in G} R_p(D)
    \end{equation}

In Berger's `source coding game' \cite{BergerSourceCodingGame}, the source is assumed to be an
adversarial player called the `switcher' in a statistical game. In this setup, the switcher is
allowed to choose any source from $G$ at any time, but must do so in a causal manner without
access to the current step's source realizations. The conclusion of
\cite{BergerSourceCodingGame} is that under this scenario,
    \begin{equation}
        R(D) = \max_{p \in \overline{G}} R_p(D)
    \end{equation}
where $\overline{G}$ is the convex hull of $G$. In his conclusion, Berger poses the question of
what happens to the rate-distortion function when the rules of the game are tilted in favor of
the switcher. Suppose that the switcher were given access to the source realizations before
having to choose the switch positions.
    The main result of this paper is that under these rules,
    \begin{equation}
        R(D) = \max_{p \in \mC} R_p(D)
    \end{equation}
where
\begin{equation} \mC = \left\{ \begin{array}{ccc}
& & \sum_{i\in \mathcal{V}} p(i) \geq \prod_{l=1}^m \sum_{i\in \mathcal{V}} p_l(i) \\
   p \in \mP & : & \forall~ \mathcal{V} \textrm{ such that } \\
 &  &  \mathcal{V} \subseteq \mX
\end{array} \right\} \label{eqn:defnC} \end{equation}

Here, the $p_l$ are the distributions of the $m$ sources and $\mP$ is the set of all
probability distributions on $\mX$.

Section \ref{sec:def} sets up the notation for the paper, and is followed by a description of
the source coding game in Section \ref{sec:game}. The main result is stated in Section
\ref{sec:mainresult}, and an example illustrating the main ideas is given in Section
\ref{sec:example}. The proofs are located in Section \ref{sec:proofs} and some concluding
remarks are made in Section \ref{sec:conclusion}.

\section{Definitions} \label{sec:def}

We work in essentially the same setup as Berger's source coding game
\cite{BergerSourceCodingGame}, and with most of the same notation. There are two finite
alphabets $\mX$ and $\mY$. Without loss of generality, $\mX = \{1,2,\ldots |\mX|\}$ is the
source alphabet and $\mY = \{1, 2, \ldots |\mY|\}$ is the reproduction alphabet. Let $\x =
(x_1, \ldots, x_n)$ denote an arbitrary vector from $\mX^n$ and $\y = (y_1, \ldots, y_n)$ an
arbitrary vector from $\mY^n$. When needed, $\x^k = (x_1, \ldots, x_k)$ will be used to denote
the first $k$ symbols in the vector $\x$.

Let $d:\mX \times \mY \rightarrow [0,\infty)$ be a distortion measure\footnote{We could allow
for infinite distortions and require that the probability that the distortion exceed
$D+\epsilon$ go to zero for all $\epsilon > 0$. The main result would hold in this setup as
well.} (any nonnegative function) on the product set $\mX \times \mY$. Then define $d_n:\mX^n
\times \mY^n \rightarrow [0,\infty)$ for $n\geq 1$ to be
    \begin{equation} d_n(\x,\y) = \frac{1}{n} \sum_{k=1}^n d(x_k,y_k)\end{equation}

Let $\mP$ be the set of probability distributions on $\mX$, $\mP_n$ the set of types of length
$n$ strings from $\mX$, and let $\mW$ be the set of probability transition matrices from $\mX$
to $\mY$ . The rate distortion function of $p \in \mP$ with respect to distortion measure $d$
is defined to be
    \begin{equation}
        R_p(D) = \min_{w \in W(p,D)} I(p,w)
    \end{equation}
where
    \begin{equation}
        W(p,D) = \B \{ w \in \mW: \sum_{i=1}^{|\mX|} \sum_{j=1}^{|\mY|} p(i)w(j|i)d(i,j) \leq D
        \B \}
    \end{equation}
and $I(p,w)$ is the mutual information\footnote{We use $\log_2$ in the report, but any base can
be used.}
    \begin{equation}
        I(p,w) = \sum_{i=1}^{|\mX|} \sum_{j=1}^{|\mY|} p(i) w(j|i)\log_2 \h[ \frac{w(j|i)}{\sum_{i' =
        1}^{|\mX|} p(i')w(j|i')}\h]
    \end{equation}
The only interesting domain of values for $R_p(D)$ is $D \in (D_{\min}(p), D_{\max}(p))$ where
    \begin{eqnarray}
        D_{\min}(p) & = & \sum_{i=1}^{|\mX|} p(i) \min_j d(i,j)\\
        D_{\max}(p) & = & \min_j \sum_{i=1}^{|\mX|} p(i)d(i,j)
    \end{eqnarray}

Let $B = \{\y_1, \ldots, \y_K\}$ be a codebook of length $n$ vectors in $\mY^n$. Define
    \begin{equation}
        d_n(\x;B) = \min_{\y \in B} d_n(\x,\y)
    \end{equation}

If $B$ is used to represent an IID source with distribution $p$, then the average distortion of
$B$ is defined to be
    \begin{equation}
        d(B) = \sum_{\x \in \mX^n} P(\x)d_n(\x;B) = E[d_n(\x;B)]
    \end{equation}
where
    \begin{equation}
        P(\x) = \prod_{k=1}^n p(x_k)
    \end{equation}

Let $K(n,D)$ be the minimum number of codewords needed in a codebook $B \subset \mY^n$ so that
$d(B) \leq D$. Then, Shannon's Rate-Distortion Theorem (\cite{ShannonRateDistortion,
WolfowitzRateDistortion}) says that if the source is IID with distribution $p$,
    \begin{equation}
        \lim_{n \to \infty} \frac{1}{n}\log_2 K(n,D) = R_p(D)
    \end{equation}

\section{The Source Coding Game} \label{sec:game}

    We suppose as in Berger's paper that a `switcher' is a player in a two person game with access
to the position of a switch which can be in one of $m$ positions. The switch position $l, 1
\leq l \leq m$ corresponds to a memoryless source with distribution $p_l(\cdot)$ that is
independent of all the other sources\footnote{There can be multiple copies of the same source.
For example, there can be any number of copies of a Bernoulli $(1/10)$ source, so long as they
are all independent. In that sense, the switcher has access to a {\em list} of $m$ sources,
rather than a set of $m$ different distributions.}. Let $\s = (s_1, s_2, \ldots, s_n)$ be the
vector of switch positions chosen by the switcher. Let $x_k$ be the switcher's output at time
$k$ and let $x_{l,k}$ be the output of the $l^{th}$ source at time $k$. When needed, $\x_l$
will denote the block of $n$ symbols for the $l^{th}$ source.

The other person in the game is called the `coder'. The coder's goal is to construct a codebook
of minimal size to ensure the average distortion between the switcher's output and
reconstruction in the codebook is at most $D$. Fix $n$ and $D\geq 0$. Let $B$ denote the
codebook chosen by the coder, and $d_n(\x;B)$ be the distortion between a vector $\x$ and the
best reproduction of $\x$ in B; in the sense of least distortion. The payoff of the game is the
average distortion, which for a particular switching strategy is
    \begin{equation} E[d(\x;B)] = \sum_{\x \in \mX^n} P_S(\x)d_n(\x;B) \end{equation}

Here $P_S(\x)$ is the probability of the switcher outputting the sequence $\x$ averaged over
any randomness the switcher chooses to use, as well as the randomness in the sources. Let
$P(\s,\x)$ be the probability of the switcher using a switching vector $\s$ and outputting a
string $\x$. Then,
    \begin{equation} P_S(\x) = \sum_{\s \in \{1, \ldots, m\}^n} P(\s,\x) \end{equation}

In Berger's original game, the coder chooses a codebook that is revealed to the switcher. The
switcher must then choose the switch position at every integer time $k$ without access to the
actual letters that the sources produce at that time. The switcher, however, has access to the
previous outputs of the switch. So in \cite{BergerSourceCodingGame}, an admissible joint
probability rule for $P(\s,\x)$ is of the form
    \begin{equation} P(\s,\x) = \prod_{k=1}^n P(s_k|\s^{k-1}, \x^{k-1})P_{s_k}(x_k) \end{equation}

    \begin{figure}
        \begin{center}
\begin{picture}(0,0)%
\includegraphics{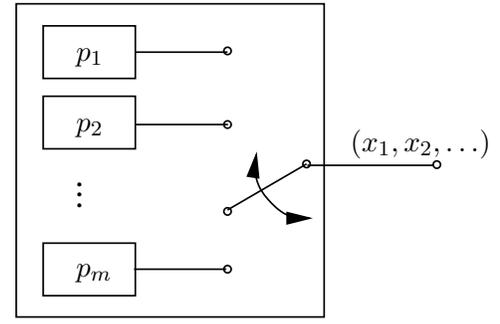}%
\end{picture}%
\setlength{\unitlength}{2763sp}%
\begingroup\makeatletter\ifx\SetFigFont\undefined%
\gdef\SetFigFont#1#2#3#4#5{%
  \reset@font\fontsize{#1}{#2pt}%
  \fontfamily{#3}\fontseries{#4}\fontshape{#5}%
  \selectfont}%
\fi\endgroup%
\begin{picture}(3844,2856)(1154,-2683)
\put(1701,-1674){\makebox(0,0)[lb]{\smash{{\SetFigFont{14}{16.8}{\rmdefault}{\mddefault}{\updefault}$\vdots$}}}}
\put(1714,-992){\makebox(0,0)[lb]{\smash{{\SetFigFont{11}{13.2}{\rmdefault}{\mddefault}{\updefault}$p_2$}}}}
\put(1714,-361){\makebox(0,0)[lb]{\smash{{\SetFigFont{11}{13.2}{\rmdefault}{\mddefault}{\updefault}$p_1$}}}}
\put(1714,-2299){\makebox(0,0)[lb]{\smash{{\SetFigFont{11}{13.2}{\rmdefault}{\mddefault}{\updefault}$p_m$}}}}
\put(4176,-1174){\makebox(0,0)[lb]{\smash{{\SetFigFont{11}{13.2}{\rmdefault}{\mddefault}{\updefault}$(x_1,
x_2, \ldots)$}}}}
\end{picture}%
        \end{center} \caption{The source coding game.}
    \end{figure}

In this discussion, we consider the case when the switcher gets to see the outputs of the $m$
sources and then has to output a letter from one of the letters that the sources produced. The
switcher outputs a letter, $x_k$, which must come from the (possibly proper) subset of $\mX$,
$\{x_{1,k}, \ldots, x_{m,k}\}$. Hence, for this `cheating' switcher, allowable strategies are
of the form
    \begin{eqnarray} \lefteqn{P(\s,\x|\x_1, \ldots, \x_m) =} \nonumber \\
    & & P(\s|\x_1,\ldots, \x_m)1(x_k = x_{s_k,k}, \1 \leq k \leq n) \end{eqnarray}
Since the sources are still IID,
    \begin{eqnarray} P(\x_1, \ldots, \x_m) = \prod_{l=1}^m \prod_{k=1}^n p_l(x_{l,k})
    \end{eqnarray}

Define the minimum number of codewords needed by the coder to guarantee average distortion $D$
as $M(n,D)$.
    \begin{equation}M(n,D) = \min\left\{ |B| : \begin{array}{c}
B \subset \mY^n, ~ E[d(\x;B)] \leq D \\
 \textrm{for all allowable} \\
 \textrm{switcher strategies }
\end{array} \right\} \end{equation}

We are interested in the exponential rate of growth of $M(n,D)$ with $n$. Define
    \begin{equation} R(D) = \lim_{n \to \infty} \frac{1}{n} \log_2 M(n,D) \end{equation}

Let $G = \{p_1(\cdot), \ldots, p_m(\cdot)\}$ be the set of $m$ distributions on $\mX$ the
switcher has access to. Let $\overline{G}$ be the convex hull of $G$. Then let
\begin{displaymath} R^\ast (D) = \max_{p \in \overline{G}} R_p(D)\end{displaymath}

The conclusion of \cite{BergerSourceCodingGame} is that $R(D)= R^\ast (D)$ when the switcher is
not allowed to witness the source realizations until committing to a switch position.

\section{Main Result} \label{sec:mainresult}

The main result is the determination of $R(D)$ in the case when the switcher gets to see the
entire block of $mn$ source outputs ahead of choosing the switching sequence.

\begin{theorem}
Let the switcher `cheat' and have access to the $n$ outputs of all $m$ sources before choosing
a symbol for each time $k$. Then,

\begin{equation} R(D) = \widetilde{R}(D) \triangleq \max_{p \in \mC} R_p(D) \end{equation}

where $\mC$ is defined in (\ref{eqn:defnC}).
\end{theorem} \vspace{.1in}

Here, we have defined $\widetilde{R}(D) = \max_{p \in \mC} R_p(D)$. The theorem's conclusion is
that when the switcher is allowed to `cheat', $R(D) = \widetilde{R}(D)$. The number of
constraints in the set $\mC$ is exponential in the size of $\mX$. Depending on the source
distributions, a large number of these constraints could be inactive. Unfortunately, $R_p(D)$
is generally not concave in $p$ for a fixed $D$, so computation of $\widetilde{R}(D)$ may not
be easy.

Qualitatively, allowing the switcher to `cheat' gives access to distributions $p \in \mC$ which
may not be $\overline{G}$. Quantitatively, the conditions placed on the distributions in $\mC$
are precisely those that restrict the switcher from producing symbols that do not occur often
enough on average. For example, let $\mathcal{V} = \{1\}$. Then for every $p \in \mC$,
    \begin{displaymath} p(1) \geq \prod_{l=1}^m p_l(1) \end{displaymath}
Since the sources are independent, $\prod_{l=1}^m p_l(1)$ is the probability that all $m$
sources produce the letter $1$ at a given time. In this case, the switcher has no option but to
output the letter $1$, hence any distribution the switcher mimics must have $p(1) \geq
\prod_{l=1}^m p_l(1)$. The same logic can be applied to all subsets $\mathcal{V}$ of $\mX$.

As commented in Section V of \cite{BergerSourceCodingGame}, $\widetilde{R}(D) = R^\ast(D)$ if
$R^\ast(D) = \max_{p \in \mP} R_p(D)$. Before giving the proof of the result, an example is
presented.

\section{An Example} \label{sec:example}

Suppose the switcher has access to two IID binary sources. Source $1$ outputs $1$ with
probability $1/3$ and source $2$ outputs $1$ with probability $1/4$. Then, since the sources
are IID across time and independent of each other, for any time $k$,
    \begin{equation}
        P(x_{1,k} = x_{2,k} = 0) = \frac{2}{3}\cdot\frac{3}{4} = \frac{1}{2}
    \end{equation}
Similarly,
    \begin{equation}
        P(x_{1,k} = x_{2,k} = 1) = \frac{1}{3}\cdot\frac{1}{4} = \frac{1}{12}
    \end{equation}
Hence,
    \begin{equation}
        P(\{x_{1,k}, x_{2,k}\} = \{0,1\}) = 1 - \frac{1}{2} - \frac{1}{12} = \frac{5}{12}
    \end{equation}

\begin{figure}
\begin{center}
\begin{picture}(0,0)%
\includegraphics{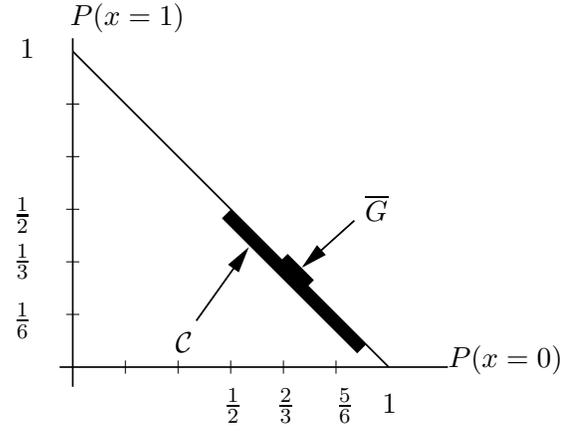}%
\end{picture}%
\setlength{\unitlength}{2763sp}%
\begingroup\makeatletter\ifx\SetFigFont\undefined%
\gdef\SetFigFont#1#2#3#4#5{%
  \reset@font\fontsize{#1}{#2pt}%
  \fontfamily{#3}\fontseries{#4}\fontshape{#5}%
  \selectfont}%
\fi\endgroup%
\begin{picture}(4923,3778)(1384,-2951)
\put(3274,-2873){\makebox(0,0)[lb]{\smash{{\SetFigFont{11}{13.2}{\rmdefault}{\mddefault}{\updefault}$\frac{1}{2}$}}}}
\put(1443,316){\makebox(0,0)[lb]{\smash{{\SetFigFont{11}{13.2}{\rmdefault}{\mddefault}{\updefault}$1$}}}}
\put(1384,-1574){\makebox(0,0)[lb]{\smash{{\SetFigFont{11}{13.2}{\rmdefault}{\mddefault}{\updefault}$\frac{1}{3}$}}}}
\put(1384,-1161){\makebox(0,0)[lb]{\smash{{\SetFigFont{11}{13.2}{\rmdefault}{\mddefault}{\updefault}$\frac{1}{2}$}}}}
\put(1384,-2106){\makebox(0,0)[lb]{\smash{{\SetFigFont{11}{13.2}{\rmdefault}{\mddefault}{\updefault}$\frac{1}{6}$}}}}
\put(4691,-2873){\makebox(0,0)[lb]{\smash{{\SetFigFont{11}{13.2}{\rmdefault}{\mddefault}{\updefault}$1$}}}}
\put(5282,-2460){\makebox(0,0)[lb]{\smash{{\SetFigFont{11}{13.2}{\rmdefault}{\mddefault}{\updefault}$P(x
= 0)$}}}}
\put(3746,-2873){\makebox(0,0)[lb]{\smash{{\SetFigFont{11}{13.2}{\rmdefault}{\mddefault}{\updefault}$\frac{2}{3}$}}}}
\put(4278,-2873){\makebox(0,0)[lb]{\smash{{\SetFigFont{11}{13.2}{\rmdefault}{\mddefault}{\updefault}$\frac{5}{6}$}}}}
\put(4536,-1145){\makebox(0,0)[lb]{\smash{{\SetFigFont{11}{13.2}{\rmdefault}{\mddefault}{\updefault}$\overline{G}$}}}}
\put(2836,-2326){\makebox(0,0)[lb]{\smash{{\SetFigFont{11}{13.2}{\rmdefault}{\mddefault}{\updefault}$\mathcal{C}$}}}}
\put(1891,616){\makebox(0,0)[lb]{\smash{{\SetFigFont{11}{13.2}{\rmdefault}{\mddefault}{\updefault}$P(x
= 1)$}}}}
\end{picture}%

\end{center}
\caption{The binary distributions the switcher can mimic. $\overline{G}$ is the set of
distributions the switcher can mimic without cheating, and $\mC$ is the set attainable with
cheating.} \label{fig:example_simplex}
\end{figure}

If at time $k$, the switcher has the option of choosing either $0$ or $1$, suppose the switcher
chooses $1$ with probability $f_1$. This strategy is memoryless, but it is an allowable
strategy for the `cheating' switcher. The coder then sees an IID binary source with a
probability of a $1$ occurring being equal to:
    \begin{equation}
        p(1) = \frac{1}{12} + \frac{5}{12} f_1
    \end{equation}
By using $f_1$ as a parameter, the switcher can produce $1$'s with a probability between $1/12$
and $1/2$. The attainable distributions are shown in Figure \ref{fig:example_simplex}. This
kind of memoryless, `conditional' switching strategy will be used for half of the proof of the
main result. If the distortion measure is Hamming distortion, clearly the switcher will choose
$f_1 = 1$ and produce a Bernoulli $1/2$ process. Regardless of the distortion measure, $\mC$
contains all the distributions on $\mX$ that the switcher can mimic.

\section{Proofs} \label{sec:proofs}
\subsection{Achievability for the coder} First, the main tool of this section is
stated.

\begin{lemma}[Type Covering \cite{BergerSourceCodingGame}] Let $\mP_n$ denote the set of types for length $n$ sequences from
$\mX$. Let $S_D(\y) \triangleq \{ \x \in \mX^n : d_n(\x,\y) \leq D\}$ be the set of $\mX^n$
strings that are within distortion $D$ of a $\mY^n$ string $\y$. Fix a $p \in \mP_n$ and an
$\epsilon > 0$. Then there exists a codebook $B = \{\y_1, \y_2, \ldots, \y_M\}$ where $M <
\exp_2(n(R_p(D) + \epsilon))$ and
    \begin{displaymath} T_p^n \subseteq \bigcup_{k=1}^M S_D(\y_k) \end{displaymath}
where $T_p^n$ is the set of $\mX^n$ strings with type $p$ for $n$ large enough.
\end{lemma}\vspace{.1in}

We now show how the coder can get arbitrarily close to $\widetilde{R}(D)$ for large enough $n$.
For $\delta > 0$, define $\mC_\delta$ as
    \begin{displaymath}
        \mC_\delta \triangleq \left\{ \begin{array}{ccc}
        & & \sum_{i\in \mathcal{V}} p(i) \geq \prod_{l=1}^m \sum_{i\in \mathcal{V}} p_l(i) - \delta \\
        p \in \mP & : & \forall~ \mathcal{V} \textrm{ such that } \\
        &  &  \mathcal{V} \subseteq \mX
        \end{array} \right\}
    \end{displaymath}

\begin{lemma}[Converse for switcher]
Let $\epsilon > 0$. For all $n$ sufficiently large \begin{displaymath}\frac{1}{n}\log_2{M(n,D)}
\leq \widetilde{R}(D) + \epsilon \end{displaymath}
\end{lemma} \vspace{.1in}

\begin{proof}
We know $R_p(D)$ is a continuous function of $p$ (\cite{CsiszarBook}). It follows then that
because $\mC_\delta$ is monotonically decreasing (as a set) with $\delta$ that for all
$\epsilon > 0$, there is a $\delta > 0$ so that
    \begin{displaymath}
        \max_{p \in \mC_\delta} R_p(D) \leq \max_{p \in \mC} R_p(D) + \epsilon/2
    \end{displaymath}

We will have the coder use a codebook such that all $\mX^n$ strings with types in $\mC_\delta$
are covered within distortion $D$. The coder can do this for large $n$ with at most $M$
codewords where
    \begin{eqnarray}
        M & < & (n+1)^{|\mX |}\exp_2(n\max_{p \in \mC_\delta} R_p(D)) \\
        & \leq & (n+1)^{|\mX|} \exp_2(n(\max_{p\in \mC} R_p(D) + \epsilon))
    \end{eqnarray}

Explicitly, this is done by taking a union of the codebooks provided by the type covering lemma
and noting that the number of types is less than $(n+1)^{|\mX|}$. Next, we will show that the
probability of the switcher being able to produce a string with a type not in $\mC_\delta$ goes
to $0$ exponentially with $n$.

Consider a type $p \in \mP_n \cap (\mP - \mC_\delta)$. By definition, there is some
$\mathcal{V} \subseteq \mX$ such that $\sum_{i\in \mathcal{V}} p(i) < \prod_{l=1}^m \sum_{i\in
\mathcal{V}} p_l(i) - \delta$. Let $\alpha_k(\mathcal{V})$ be the indicator function
    \begin{displaymath} \alpha_k(\mathcal{V}) = \prod_{l=1}^m \1(x_{l,k} \in \mathcal{V}) \end{displaymath}
$\alpha_k$ indicates the event that the switcher cannot output a symbol outside of
$\mathcal{V}$ at time $k$. Then $\alpha_k(\mathcal{V})$ is a Bernoulli random variable with a
probability of being $1$ equal to $Q(\mathcal{V}) \triangleq \prod_{l=1}^m \sum_{i\in
\mathcal{V}} p_l(i)$. That is, we can envision $\alpha_k(\mathcal{V})$ as being a sequence of
IID binary random variables with distribution $q' \triangleq (1-Q(\mathcal{V}),
Q(\mathcal{V}))$.

Now for our type $p \in \mP_n \cap (\mP - \mC_\delta)$, we have that for all strings $\x$ in
the type class $T_p$, $\frac{1}{n} \sum_{i=1}^n \1(x_i \in \mathcal{V}) < Q(\mathcal{V}) -
\delta$. Let $p'$ be the binary distribution $(1 - Q(\mathcal{V}) + \delta, Q(\mathcal{V}) -
\delta)$, assuming $\delta$ is small enough to make this a distribution (if not, make delta
small enough). Therefore $||p' - q'||_1 = 2\delta$, and hence $D(p'||q') \geq \delta/\ln 2$ by
Pinsker's inequality. Using standard types properties \cite{CoverBook} gives
    \begin{eqnarray*}
        P\bigg ( \frac{1}{n} \sum_{k=1}^n \alpha_k(\mathcal{V}) < Q(\mathcal{V}) - \delta \bigg) & \leq & \exp_2(-nD(p'||q')) \\
        & \leq & \exp_2(-n\delta/\ln2)
    \end{eqnarray*}

If we let $E$ be the event that $\x$ has a type which is not in $\mC_\delta$, we just sum over
types not in $\mC_\delta$ to get
    \begin{eqnarray*}
        P(E)& \leq & \sum_{p \in \mP_n \cap (\mP - \mC_\delta)} \exp_2(-n\delta/\ln 2) \\
        & \leq & (n+1)^{|\mX|}  \exp_2(-n\delta/\ln 2) \\
        & = & \exp_2\l(-n \bigg( \frac{\delta}{\ln 2} - |\mX| \frac{\ln (n+1)}{n}\bigg)\r)
    \end{eqnarray*}

Now let $d^\ast = \max_{x,y} d(x,y) < \infty$. Then, regardless of the switcher strategy,
    \begin{displaymath}
        E[d(\x;B)] \leq D + d^\ast \cdot \exp_2 \Bigg(-n \bigg( \frac{\delta}{\ln 2} - |\mX| \frac{\ln
        (n+1)}{n}\bigg)\Bigg)
    \end{displaymath}

So for large $n$ we can get arbitrarily close to distortion $D$ while the rate is at most
$\widetilde{R}(D) + \epsilon$. Using the fact that the rate-distortion function is continuous
in $D$ gives us that the coder can achieve at most distortion $D$ on average while the rate is
at most $\widetilde{R}(D) + \epsilon$. Since $\epsilon$ is arbitrary, $R(D) \leq
\widetilde{R}(D)$.
\end{proof}

\subsection{Achievability for the switcher} This section considers why
$R(D) \geq \widetilde{R}(D)$. We will show that the switcher can target any distribution $p \in
\mC$ and produce a sequence of IID symbols with distribution $p$. In particular, the switcher
can target the distribution that yields $\max_{p \in \mC} R_p(D)$ and Shannon's rate distortion
theorem gives $R(D) \geq \widetilde{R}(D)$.

The switcher will use a memoryless randomized strategy. Let $\mathcal{V} \subseteq \mX$ and
suppose that at some time $k$ the set of symbols available to choose from for the switcher is
exactly $\mathcal{V}$. That is $\{x_{1,k}, \ldots, x_{m,k}\} = \mathcal{V}$. Define
$\beta(\mathcal{V}) \triangleq P(\{x_{1,1}, \ldots, x_{m,1}\} = \mathcal{V})$ to be the
probability that at any time the switcher can choose any element of $\mathcal{V}$ and no other
symbols. Then let $f(i|\mathcal{V})$ be a probability distribution on $\mX$ with support
$\mathcal{V}$, i.e. $f(i|\mathcal{V}) \geq 0,~ \forall  ~i \in \mX$, $f(i|\mathcal{V}) = 0$ if
$i \notin \mathcal{V}$, and $\sum_{i\in \mathcal{V}} f(i|\mathcal{V}) = 1$. The switcher will
have such a randomized rule for every nonempty subset $\mathcal{V}$ of $\mX$ such that
$|\mathcal{V}| \leq m$. Let $\mD$ be the set of distributions on $\mX$ that can be achieved
with these kinds of rules, so
    \begin{displaymath}
        \mD \triangleq \left\{ \begin{array}{ccc}
        & & p(\cdot) = \sum_{\mathcal{V} \subseteq \mX, |\mathcal{V}|\leq m} \beta(\mathcal{V}) f(\cdot|\mathcal{V}), \\
         p \in \mP & : & \forall~ \mathcal{V} \textrm{ s.t. } \mathcal{V} \subseteq \mX, ~ |\mathcal{V}| \leq m, \\
        &  &  f(\cdot|\mathcal{V}) \textrm{ is a PMF on} ~\mathcal{V} \end{array} \right\}
    \end{displaymath}

It is clear from the construction of $\mD$ that $\mD \subseteq \mC$ because the conditions in
$\mC$ are those that prevent the switcher only from producing symbols that do not occur enough,
but put no further restrictions on the switcher. So we need only show that $\mC \subseteq \mD$.
The following gives such a proof by contradiction.

\begin{lemma}[Achievability for switcher]
The set relation $\mC \subseteq \mD$ is true.
\end{lemma}\vspace{.1in}
\begin{proof}
Suppose $p \in \mC$ but $p \notin \mD$. It is clear that $\mD$ is a convex set. Let us view the
probability simplex in $\mathbb{R}^{|\mX|}$. Since $\mD$ is a convex set, there is a hyperplane
through $p$ that does not intersect $\mD$. Hence, there is a vector $(a_1, \ldots, a_{|\mX|})$
such that $\sum_{i=1}^{|\mX|} a_i p(i) = t$ for some real $t$ but $t < \min_{q \in \mC}
\sum_{i=1}^{|\mX|} a_i q(i)$. Without loss of generality, assume $a_1 \geq a_2 \geq \ldots \geq
a_{|\mX|}$ (otherwise permute symbols). Now, we will construct $f(\cdot| \mathcal{V})$ so that
the resulting $q$ has $\sum_{i=1}^{|\mX|} a_i p(i) \geq \sum_{i=1}^{|\mX|} a_i q(i)$, which
contradicts the initial assumption. Let
    \begin{displaymath}
        f(i|\mathcal{V}) \triangleq \left\{ \begin{array}{ccc} 1 & &~ \textrm{ if $i = \max(\mathcal{V})$}\\
        \\
        0 & & \textrm{else} \end{array} \right.
    \end{displaymath}

For example, if $\mathcal{V} = \{1, 5, 6, 9\}$, then $f(9|\mathcal{V}) = 1$ and
$f(i|\mathcal{V}) = 0$ if $i \neq 9$. Call $q$ the distribution on $\mX$ induced by this choice
of $f(\cdot|\mathcal{V})$. Recall that $Q(\mathcal{V}) = \prod_{l=1}^m \sum_{i \in \mathcal{V}}
p_l(i)$. Then, we have
    \begin{eqnarray*}
        \sum_{i=1}^{|\mX|} a_i q(i) & = & a_1 Q(\{1\}) + a_2 [Q(\{1,2\} - Q(\{1\})] + \\
        & \hspace{-.5in}\cdots & \hspace{-0.25in}+ a_{|\mX|} [Q(\{1,\ldots, |\mX|\}) - Q(\{1,\ldots, |\mX| -
        1\})]
    \end{eqnarray*}

By the constraints in the definition of $\mC$, we have the following inequalities for $p$:
    \begin{eqnarray*}
        p(1) & \geq & Q(\{1\}) = q(1) \\
        p(1) + p(2) & \geq & Q(\{1,2\}) = q(1) + q(2) \\
        & \vdots &\\
        \sum_{i=1}^{|\mX|-1} p(i) & \geq & Q(\{1,\ldots, |\mX|-1\}) = \sum_{i=1}^{|\mX| -1} q(i)
    \end{eqnarray*}

Therefore, the difference of the objective is
    \begin{eqnarray*}
         \lefteqn{\sum_{i=1}^{|\mX|} a_i(p(i) - q(i)) =} & &\\
         & & a_{|\mX|}\bigg[\sum_{i=1}^{|\mX|} p(i) - q(i)\bigg] + \\
         & & (a_{|\mX|-1} - a_{|\mX|})\bigg[\sum_{i=1}^{|\mX|-1} p(i) - q(i)\bigg] + \\
        & & \cdots + (a_1-a_2)\bigg[p(1) - q(1)\bigg] \\
        & = & \sum_{i=1}^{|\mX| - 1} (a_i - a_{i+1})\bigg[ \sum_{j=1}^i p(j) - \sum_{j=1}^i q(j)
        \bigg]\\
        & \geq & 0
    \end{eqnarray*}

The last step is true because of the monotonicity in the $a_i$ and the inequalities we derived
earlier. Therefore, we see that $\sum_{i=1}^{|\mX|} a_ip(i) \geq \sum_{i=1}^{|\mX|} a_iq(i)$
for the $p$ we had chosen at the beginning of the proof. This contradicts the assumption that
$\sum_{i=1}^{|\mX|} a_ip(i) < \min_{q \in \mD} \sum_{i=1}^{|\mX|} a_iq(i)$, therefore it must
be that $\mC \subseteq \mD$.
\end{proof}

\section{Conclusion} \label{sec:conclusion}
    The rate-distortion function for the `cheating' switcher has been described. It is the maximization
of the IID rate-distortion function over the distributions the switcher can simulate. It was
assumed the switcher had access to all source outputs ahead of time, but the proof required
only that the switcher had access to the source realizations for one step ahead at each time.

In this paper, the sources were independent and memoryless. A minor tweak to the argument also
gets the rate-distortion function if the sources are dependent but still memoryless. The region
$\mC$ would just be modified to become:
\begin{displaymath} \mC = \left\{ \begin{array}{ccc}
& & \sum_{i\in \mathcal{V}} p(i) \geq P\big(\cup_{l=1}^m x_{l,1} \subset \mathcal{V} \big) \\
   p \in \mP & : & \forall~ \mathcal{V} \textrm{ such that } \\
 &  &  \mathcal{V} \subseteq \mX
\end{array} \right\} \end{displaymath}

A more interesting problem is to consider what happens when the sources are independent but
have memory. Apparently, Dobrushin \cite{DobrushinMemory} has analyzed the case of the
non-cheating switcher with independent sources with memory. One could imagine that, perhaps,
giving the switcher access to all source realizations could result in the ability to simulate
memoryless sources from a collection of sources with memory.

Similar techniques might also prove useful in considering a cheating `jammer' for an
arbitrarily varying channel. While the problem is mathematically well defined, it seems
unphysical in the usual context of jamming or channel noise. The idea may make more sense in
the context of watermarking, where the adversary can try many different attacks on different
letters of the input before deciding to choose one for each.

\section*{Acknowledgment}
The authors would like to thank the NSFGRFP for partial support of this research. Also, we
thank Prof. Michael Gastpar, the students of the Fall 2006 EE290S course at UC Berkeley, and
the reviewers for helping to refine the presentation of this work.

\bibliographystyle{IEEEtran}

\end{document}